# A New Drop Fluidics Enabled by Magnetic Field Mediated Elasto-Capillary Transduction

Saheli Biswas[1], Yves Pomeau[2] and Manoj K. Chaudhury[1*]

[1]*Department of Chemical and Biomolecular Engineering, Lehigh University, Bethlehem Pennsylvania 18015 USA*

[2]*University of Arizona, Department of Mathematics, Tucson, USA*

**Abstract.** This research introduces a new drop fluidics, which uses a deformable and stretchable elastomeric film as the platform, instead of the commonly used rigid supports. Such a soft film impregnated with magnetic particles can be modulated with an external electromagnetic field that produces a vast array of topographical landscapes with varying surface curvature, which, in conjunction with capillarity, can direct and control motion of water droplets efficiently and accurately. When a thin layer of oil is present on this film that is deformed locally, a centrosymmetric wedge is formed. A water droplet placed on this oil laden film becomes asymmetrically deformed thus producing a gradient of Laplace pressure within the droplet setting it to motion. A simple theory is presented that accounts for the droplet speed in terms of such geometric variables as the volume of the droplet and the thickness of the oil film covering the soft elastomeric film, as well as such material variables as the viscosity of the oil and interfacial tension of the oil-water interfaces. Following the verification of the theoretical result using well-controlled model systems, we demonstrate how the electromagnetically controlled elasto-capillary force can be used to manipulate the motion of single and/or multiple droplets on the surface of the elastomeric film and how such elementary operations as drop fusion and thermally addressed chemical transformation can be carried out in aqueous droplets. It is expected that the resulting drop fluidics would be suitable for digital control of drop motion by simply switching on and off the electromagnetic fields applied at different positions underneath the elastomeric film. We anticipate that this method of directing and manipulating water droplets is poised for its applications in various biochemical reaction engineering, an example of which is Polymerase Chain Reaction (PCR).

* To whom correspondence should be addressed

## INTRODUCTION

Control of the motion[1-4] of small amounts of liquids at meso and macro scales is fundamental to the development of various types of fluidic devices that are of importance in both thermal and water management technologies. In conventional microfluidics[5-9], flow is controlled in interconnected channels for which an external pressure is applied. By contrast, in various types of discrete flow control [1-4, 10-15], a curvature gradient is imposed in a droplet that in turn induces a Laplace pressure gradient in it, thus causing it to move. The velocity of the drop is dictated from the interplay of an unbalanced Laplace force and the viscous drag, so that it is proportional to the capillary velocity (ratio of surface tension to viscosity) of the drop. When the viscous force is insignificant, the excess surface free energy of the drop can be converted to its kinetic energy, which has given rise to a plethora of reports related to the jumping[16-20] of droplets. Drop propulsion mediated either by electro wetting[21, 22] or dielectrophoretic[23] forces have also been effective in several drop transport processes. A unique situation arises when drops growing on a surface due to the condensation of vapor coalesce with each other, in which the coalesced drop exhibits a diffusive[24] motion. It was found recently[25] that when an absorbing boundary is present to remove these diffusive droplets, a long range directed motion of drops can be generated without the use of a continuous gradient of surface energy. One significant impediment of the droplet motion, however, is defects that pin the contact line of a drop. However, either an external vibration[26, 27] or coalescence[3] of several drops have been found to overcome the pinning forces. On a surface with finite wetting hysteresis ratcheting motion[27-34] of drops can also be created either by using an asymmetry of surface in conjunction with an external symmetric vibration or with a homogeneous surface assisted by an asymmetric vibration[32, 33]. In spite of these developments, drop fluidic technologies require extensive instrumentations that increase its cost and/or it suffers from the drawback that the surface becomes eventually contaminated that prevents its repeated use. What would be ideal is a drop fluidic device that would be inexpensive to manufacture, the speed of operation would be relatively fast, and its surface would be free from contamination so that it can be used repeatedly.

In a drop fluidic device, it is virtually impossible to avoid contamination if the drop makes contact with the substrate. In certain cases, it might be possible to avoid contamination if the components present in a drop do not preferentially adsorb on the supporting surface, as we observed previously[33] with a drop containing silicone glycol surfactant. In some cases, it has been possible to immerse the droplet in oil[23], which avoids contamination. However, the mechanisms underlying such a drop propulsion require elaborate micro-fabrication of the substrate. Leidenfrost effect induced transport of drop[35], in which a vapor layer separates the drop from the substrate, may also be sought to solve the contamination issue; however, such a method is usually suitable for high temperature applications, which has been exemplified in the synthesis of nanoparticles for which the thermal condition in the drop is exactly what is needed[36]. Precise methods, nevertheless, need to be developed to stop and actuate a drop and to do so at low temperature[37] before such a method finds wider applications. Recently, some promising developments[38-40] in actuating drops via magnetic field have also taken place, by impregnating magnetic particles in an oil with which water drops exhibit very low hysteresis, or coating the drop with magnetic particles yielding similar effect.

The platform used in all the drop fluidic devices is, however, static, the efficacy of which is ultimately limited by the spatial density of the complex array of electronic circuitries. In order to circumvent such a limitation, we were inspired by a question: what if the platform of the drop fluidic device is not static but it is one in which the landscape can dynamically evolve? To illustrate this point, let us consider a platform made of elastically deformable film embedded with magnetic particles arranged in a square grid, where each or few of the particle(s) can be independently addressed by an external magnetic field. As with such a square matrix array of particles, the local deformations of the film are amenable to a Boolean on/off sequence, an extremely large array of curvature induced energy landscapes can be produced on the film. If, now, a mechanism is designed with which a liquid drop would sense such a landscape and be disposed to movement, vast possibilities of the direction of motion of a single, binary and multiple drops could be designed on the surface.

What we describe here is our first set of results demonstrating the feasibility of the above concept by using a magnetically addressable elasto-capillary effect that results from the asymmetric deformation of the water droplet on the surface of an oil laden elastomeric film. The interfacial thermodynamic conditions of all the interfaces here are such that the water drop does not adhere to the elastomer. The oil also spreads completely on the surface of water, thereby ensuring that the water drop is softly confined between the air-oil and solid-oil interfaces. As soon as the elastomeric film is deformed to a circular wedge, the drop becomes subjected to an unbalanced Laplace force prompting it to move towards the center of the deformed region where the unbalanced capillary force vanishes. Now by changing the landscape of the film, the direction of the motion of the drop can be diverted, thereby creating enumerable gaits of droplet motion.

## RESULTS AND DISCUSSIONS

### Mechanism of Droplet Propulsion

As indicated above, the fundamental method of drop propulsion is to create a differential curvature on two opposite sides of a drop that propels it towards the direction where the curvature differential would be minimal. Such a curvature gradient can be produced with a surface displaying a gradient of wettability[1], placing the drop on a conical wire[41, 42] or in a geometrical wedge[12]. The physics underlying all these droplet propulsion mechanisms has been extensively reviewed in recent literature[1, 12]. Our starting point is also to create a curvature differential in a drop placed in a wedge[12], but in which one of the surfaces is a deformable air-oil interface.

Let us consider a water drop interacting with a solid surface, through a third phase, which in our case is an immiscible oil. The stability of the thin oil film intervening the water drop and the solid support is determined by the long range electrodynamic fluctuation forces. While the electrodynamic fluctuations at all possible spectral ranges contribute to interactions in condensed phases, two significant contributors to the net interaction arise from the excitations of the media in the ultraviolet and microwave frequencies. If the dielectric susceptibilities of the three phases are either in ascending or descending order (i.e. $\varepsilon_s > \varepsilon_o > \varepsilon_w$ or $\varepsilon_s < \varepsilon_o < \varepsilon_w$) at all frequencies, a repulsion is ensured between water drop and the substrate, so the oil film remains stable against all perturbations. Although, it is quite possible to select a solid substrate and an oil based on their electrodynamic spectra to ensure such a stability of the oil film, the following simple thermodynamic method suffices for that purpose[43]. We begin by writing down the free energy of adhesion of the water drop and the substrate through oil using Dupre's equation:

$$\Delta G = \gamma_{sw} - \gamma_{ow} - \gamma_{so} \tag{1}$$

where, $\gamma_{sw}$, $\gamma_{ow}$ and $\gamma_{so}$ are the interfacial tensions of solid-water, oil-water and solid-oil respectively. If the above interfacial tensions are such that $\gamma_{sw} > \gamma_{ow} + \gamma_{so}$, $\Delta G$ would be greater than zero; consequently, the water drop will not adhere to the solid.

The interfacial tension of the oil-water interfaces (used in this experiment) as estimated using a drop weight method is $\gamma_{ow}$ ~25 mN/m, whereas the interfacial tension of the solid water interface is $\gamma_{sw}$ ~ 40 mN/m. Since the oil interacts with the solid via dispersion interaction, $\gamma_{so}$ ~ 1 mN/m. With these values of interfacial tensions, the condition that $\Delta G > 0$ is readily satisfied ensuring that a thin oil film remains stable between a drop of water and the substrate, which in our case is either a Trimethylsiloxy Terminated Polydimethylsiloxane (DMS-T22) grafted silicon wafer or a Polydimethylsiloxane (PDMS) elastomer. This non-adherence of the droplet to the substrate is also beneficial when it is desired that the droplet is removed from the oil laden substrate at any stage of a specific drop fluidic operation. As shown in Figure 1, the entire drop can be removed into a capillary tube without any trace of it remaining on the substrate, which would be difficult to perform if the drop adheres to the substrate. In this latter case, a smaller droplet would remain behind the substrate as a result of the pinch off instability of the parent droplet during the detachment process. In addition to the fact that the drop does not adhere to the substrate, another fortunate situation with all the oils used in our experiments is that they spread readily on the surface of water. Taken together, we have a situation in which a water droplet remains sandwiched between the air-oil and the solid-oil interfaces. The Laplace pressure exerted by the film, the tension of which is composed of the interfacial tensions of the air/oil and oil/water interfaces, press the water drop downward, which is counteracted by an equivalent Laplace force generated within the drop due to its own deformation.

**A drop under soft confinement**

A droplet sandwiched between air and the solid (Figure 1) deforms against the solid support with a contact width *b,* the value of which is determined by the balance between the total Laplace pressure exerted by the thin film above the drop and the spring action of the drop itself. In order to make an estimate of $b$, we draw inspiration from the Hertzian contact mechanics[44], according to which, the width of the contact deformation of a sphere (radius $R$ and modulus $E$) on a flat surface due to a normal force $F_N$ is $b^3 \sim F_N R/E$. For our problem to be dimensionally similar to that of Hertz, the normal force is to be expressed as the total vertical Laplace force: $F_N \sim \gamma_F a^2/R$ with $a^2 = 2Rh - h^2$ that deforms the water drop having an equivalent modulus as $E \to \gamma_{ow}/R$. $\gamma_F$ is the film tension that is contributed by the air/oil and oil/water interfaces. All the geometrical variables of this problem are shown in Figure 1. We thus obtain the following expression for $b$:

$$b \sim \left[(\gamma_F/\gamma_{ow})R(2Rh - h^2)\right]^{1/3} \qquad (2)$$

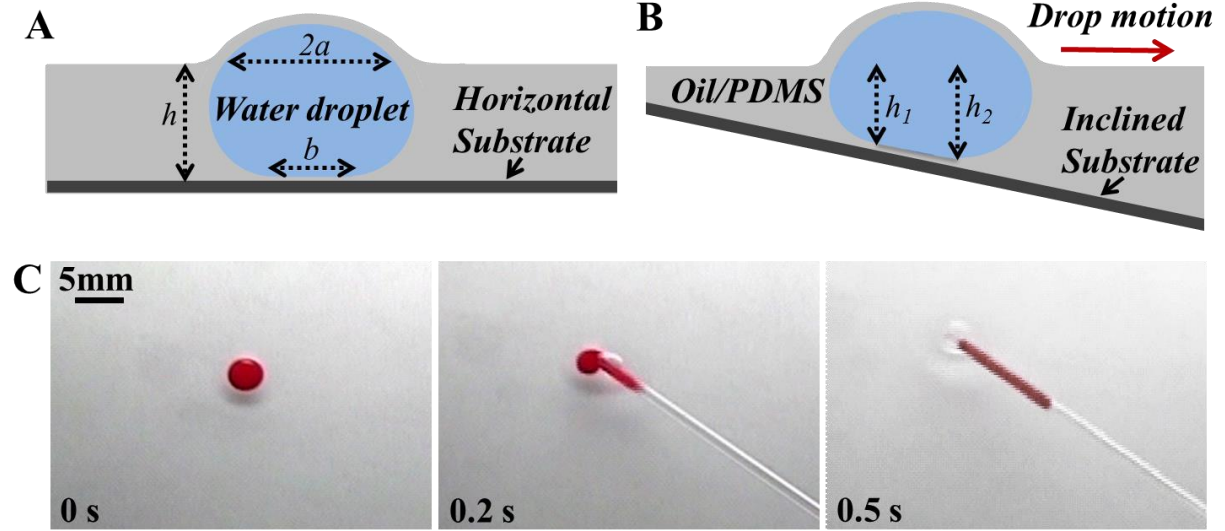

**Figure 1.** A) Schematic of a water droplet on a DMS-T22 grafted silicon wafer through a film of oil in which the water drop does not contact with the support. The oil spreads on water drop as well, causing the drop to be deformed by the Laplace pressure generated by the curved film above the droplet B) when the support is tilted, the left-right symmetry of the Laplace pressure (as in figure A) is broken. A gradient of this pressure propels the drop from the thinner towards the thicker part of the oil film C) A water drop sandwiched between the air-oil interface (a poly phenyl methyl functional silicone oil ) and a solid substrate can be removed by capillary action in just 0.3 seconds. The region in the oil, from which the water drop was recovered, shows an indentation upon the removal of the drop as it has not fully relaxed due to viscous slowing down.

At this juncture, a question might be asked regarding the importance of the role of gravity in the deformation of the water drop. The explicit condition for neglecting gravity is that the drop must be smaller than the capillary length and that this length becomes bigger as the density difference gets smaller. The effect of gravity can thus be reduced by using oils of specific gravities (0.97 to 1.05) close to that of water. As most part of the water drop is immersed in oil, it experiences negligible effect of gravity because of very small density contrast. However, the part of the drop above the base of the air/oil interfaces may, conceivably, be deformed by gravity. Thankfully, however, the film tension that prevents this gravity induced deformation is composed of two interfaces (oil/water and air/oil). In any event, we measured the radius of curvature of the spherical cap of the drop above the base of the oil, and found it to be very close (see Supporting Information, Figure S1) to that of the deposited droplet (11$\mu$L to 55 $\mu$L) in spherical shape approximation. These observations thus strengthen the assumption underlying the derivation of eq 2, in that the deformation of the drop against the rigid support is mainly due to Laplace force exerted by the spherical cap.

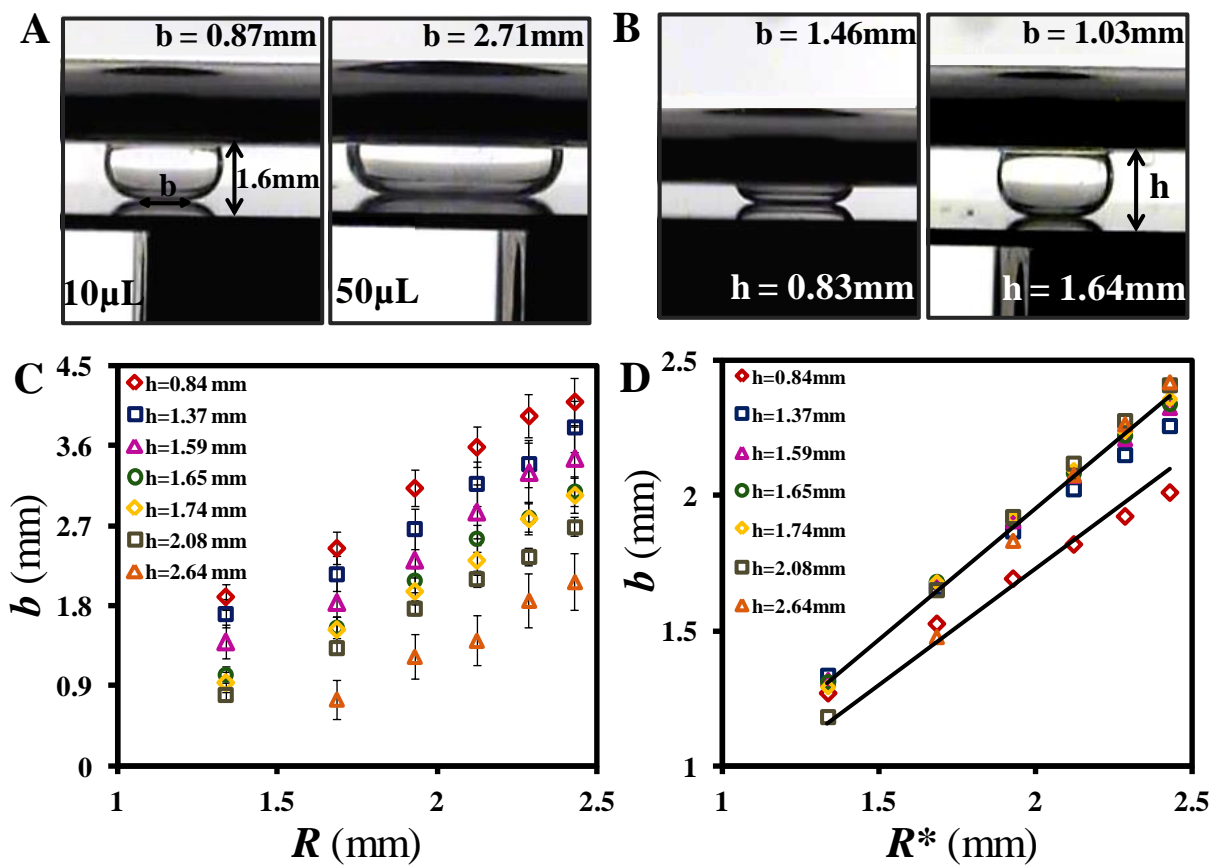

**Figure 2.** Side views of water drops sandwiched between a PDMS grafted silicon wafer and an air-silicone oil interface show that the width of the base (*b*) contact diameter increases with the size of the drop (A), but it decreases with the thickness *h* of the oil above the drop (B). The quantitative data are summarized in figure C. D. The data presented in figure C are processed according to eq 2 by plotting the width (*b*) of contact as a function of $R^* = (R(2Rh - h^2))^{1/3}$.

In order to verify the validity of eq 2, we used a polydimethylsiloxane grafted (5 nm thick) silicon wafer as the substrate that was covered with a silicone oil of density 0.95. Water drops of volume ranging from 3 $\mu$L to 70 $\mu$L were deposited on the flat silicon wafer by varying the thickness (*h*) of the oil between air and the substrate. The results summarized in Figure 2 show that the width of the contact deformation of the drop and the substrate decreases with *h*, but increases with *R* as is predicted by eq 2 qualitatively. When all the data are rescaled by defining an equivalent drop radius as $R^* = (R(2Rh - h^2))^{1/3}$, all the *b-R\** data (except for the thinnest film) cluster around each other that is in compliance with eq 2. Even though the thinnest film deviated somewhat from the rest, we point out that the slope of the *b-R\** plot differs from the rest by only 10%.

**Motion of a Squeezed Drop in a Wedge Geometry.**

Let us now turn our attention to the case when the silicon wafer is placed at an angle with respect to the air-oil interface. In this case, the drop is deformed asymmetrically that, in turn, exerts an unbalanced Laplace force to the drop. The Laplace pressure difference between the left ($P_1$) and right ($P_2$) sides of the droplet is approximately given by the following equation:

$$P_1 - P_2 \approx 2\gamma_{ow}\left(\frac{1}{h_1} - \frac{1}{h_2}\right) \qquad (3)$$

here, $h_1$ and $h_2$ are the depths of the oil layer on either side of the drop as shown in Figure 1. As the thickness *h* varies linearly with distance *x*, eq 3 can also be written as follows:

$$(P_1 - P_2)n \approx \gamma_{ow}\nabla h(b/h^2) \qquad (4)$$

here, *n* is the unit vector. This pressure acts on the area projected by the drop normal to the direction of motion, thus the total force experienced by the drop is $\sim (P_1-P_2)R^2$, or

$$\vec{F} \sim \gamma_{ow}\nabla h(bR^2/h^2) \qquad (5)$$

Substitution of the expression for $b$ (eq 2) in eq 5 leads to an expression for the total capillary motor force, which can be equated to a Stokes like drag force $\sim \alpha\eta R V$ to obtain the velocity ($V$) of the drop. We ignore any contact line drag[45], as the drop subtends a 180° contact angle with the substrate. Here $\eta$ is the viscosity of the oil and the numerical value of $\alpha$ is $4\pi$ according to the Hadamard-Rybczynski analysis[46, 47] for a fully immersed water drop moving through an oil with a large viscosity contrast. Although the numerical value of $\alpha$ is likely to differ from $4\pi$ in our case, it does not prevent us from writing a scaling law relating the drop velocity in terms of the capillary velocity ($V^* = \gamma_{ow}/\eta$) as follows:

$$\vec{V} \approx V^* \nabla h (\gamma_F/\gamma_W)^{1/3} \left[R(2Rh - h^2)\right]^{1/3} R/h^2 \qquad (6)$$

Eq 6 predicts that the droplet velocity increases with $R$ and decreases with $h$ with exponents greater than unity; both these predictions have been verified experimentally (see below).

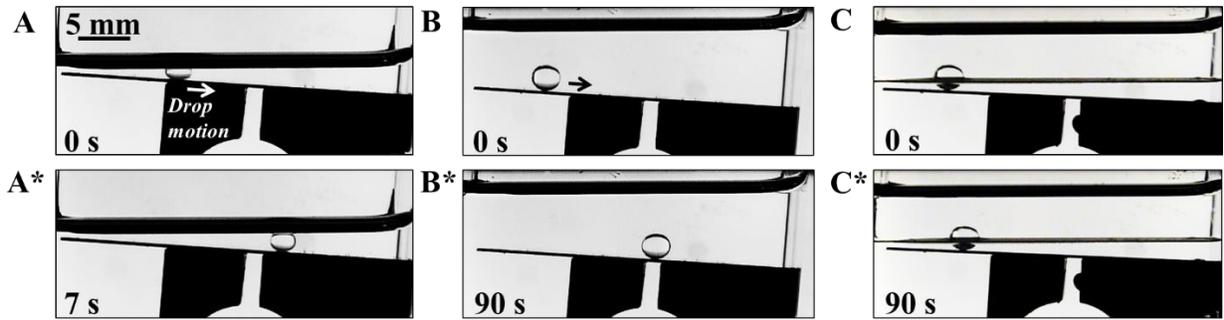

**Figure 3.** A). A 10 μl size water drop sandwiched between a PDMS grafted silicon wafer and an air-oil (DMS-T12) interface moves right (see up and down panels, e.g. A to A*, etc). B). By contrast, even a larger (20 μl) water drop moves very slowly when it is completely submerged in the oil. C). The importance of the Laplace pressure induced compression of a water drop is demonstrated here using two oil layers (DMS-T12 on top of a high temperature silicone oil), which are immiscible with each other (Note: high temperature silicone oil is the commercial name of the oil meaning it is stable at high temperature). Here, even though the drop is in between the silicon wafer and the interface produced by two different oils, the interfacial tension of the oil is negligible compared to the surface tension of an air-oil interface. Because of very low Laplace pressure, a 20 μl size water drop moves much more slowly compared to that shown in Figure 3A. The wedge angle is 4° in all the figures.

Figure 3 summarizes the basic observations of the motion of a water droplet placed at the wedge formed by the air-oil and solid-oil interfaces. The solid support used here is a silicon wafer, which is modified by grafting an ultra-thin (5 nm) film of polydimethylsiloxane, the surface energy of which is ~ 24 mN/m. A rectangular quartz cell was partially filled with a silicone oil, following which the PDMS grafted silicon wafer was placed inside it on the top of a solid support. The quartz cell containing the silicon wafer was rotated about two orthogonal axes in such a way that it was as flat as possible in one direction, but was slightly tilted in the orthogonal direction. In these experiments, the air-oil interface remains always flat, while the wedge angle is modified by tilting the silicon wafer relative to the air-oil interface. As a drop of water is placed atop the silicon wafer through oil, it moves towards the wider gap of the wedge (Movie1). Gravity has negligible effect in this type of droplet motion that can be easily shown by an experiment in which it moves very slowly when it is completely immersed in the oil (Movie2). By contrast, the importance of the surface tension of the oil, and thus the role of the resulting Laplace pressure, becomes evident in an experiment in which part of the water drop protrudes out of a thin oil film, the surface tension of which is reduced by placing another immiscible oil above it (Figure 3). The plan view of the motion of a water droplet over the silicon wafer covered with a thin oil film is shown in Figure 4 (Movie3). It is interesting to note that when more than one droplet are deposited on the silicon wafer, they attract each other due to the ensuing capillary interaction[48, 49] between them, often fusing before they reach the end of the wedge (Movie4). This fusion of droplets in the thin oil film is of significant importance in designing a suitable drop fluidic device to carry out specific reactions inside the droplets.

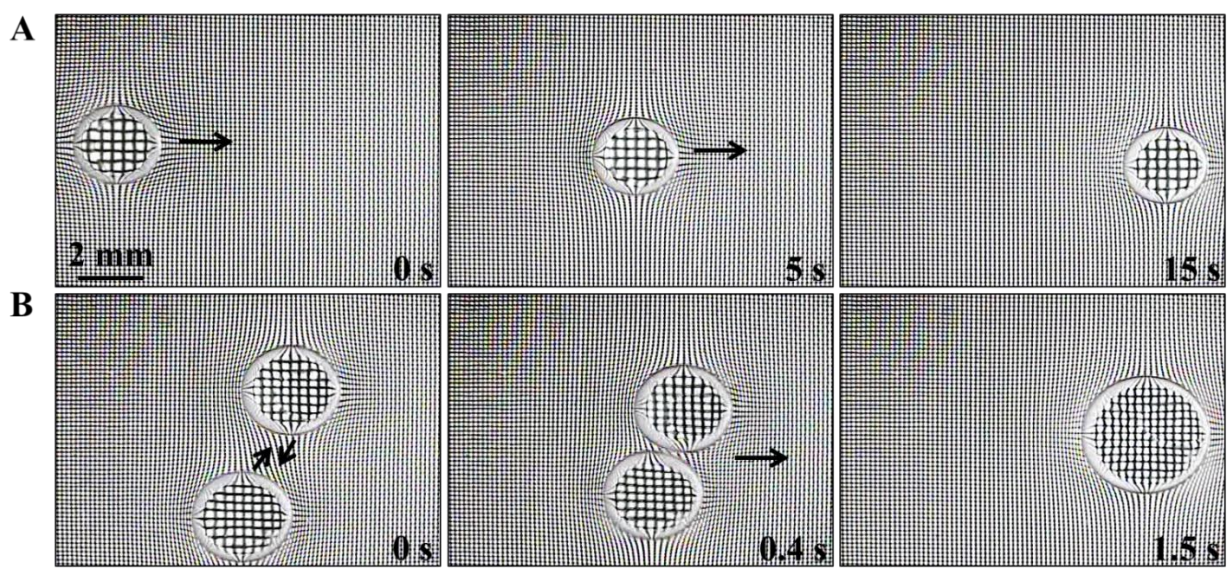

**Figure 4.** Plan views of the motion of a water droplet moving on an inclined substrate created by a capillary wedge. The upper panel (A) shows the motion of a single droplet from left to right. (B) When more than one droplet moves via capillary wedge, they attract each other via capillary forces created by the surface deformations of the oil, often fusing with each other as they move. The contrast between the water drop and the oil is enhanced by placing a metal grid below the test cell. The wedge angle is 1° in all the figures.

**Table1**. **Density and Viscosity of the test liquids (see also experimental section)**

| *Liquid* | $\rho$ (gm/cc) | $\eta$ (cSt) | *Liquid* | $\rho$ (gm/cc) | $\eta$ (cSt) |
|---|---|---|---|---|---|
| AR 20 (A) | 1.01 | 23.3 | 1:1 A/B (E) | 1.03 | 61.8 |
| High temperature silicone oil (B) | 1.05 | 183.3 | 1:3 A/B (F) | 1.04 | 105.7 |
|  |  |  | DMS-T22 (G) | 0.968 | 200 |
| 9:1 A/B (C) | 1.01 | 28.3 | DMS-T31 (H) | 0.971 | 1000 |
| 3:1 A/B (D) | 1.02 | 35 | 1:1 DMS-T22/DMS-T32 (J) | 0.969 | 485 |

**Note**: "High temperature silicone oil" is the commercial name of this oil meaning it is stable at high temperature. Chemically, both this oil and AR 20 are phenyl methyl siloxanes. Both DMS-T22 and DMS-T31 are dimethyl siloxane oils. These oils when mixed at different proportions, as indicted in the table, produce liquids of different viscosities.

The drop propulsion experiments were carried out by varying $R$, $h$, $\theta$ and $V^*$. The results of the experiments carried out at a constant $\theta$ and $V^*$ show that the drop velocity varies non-linearly with both $R$ and $h$ as can be gleaned from eq 6. When these velocities are plotted against the dimensionless radius $RR^*/h^2$, all the data (Figure 5) reasonably cluster around a straight line (correlation coefficient 98%) passing through origin (0, 0), which vindicates the geometrical factor of eq 6. Since the velocity depends so strongly on both $R$ and $h$, extreme care needs to be taken in quantifying the motion of the droplet in a wedge, where $h$ changes along its length, which was also observed by Reyssat[12] for bubbles moving in a wedge of two rigid walls.

Thus, when the wedge angle is altered, we found it convenient to measure the velocity in the middle of the wedge in order to provide a uniformity of results obtained from different experiments. Thus by keeping both $R$ and $h$ as constants, the velocity of water drop was measured at different wedge angles and with liquids of various viscosities that, in turn, alters $V^*$. The function $(\gamma_F/\gamma_W)^{1/3}$ provides a second order effect and remains more or less constant in our experiments.

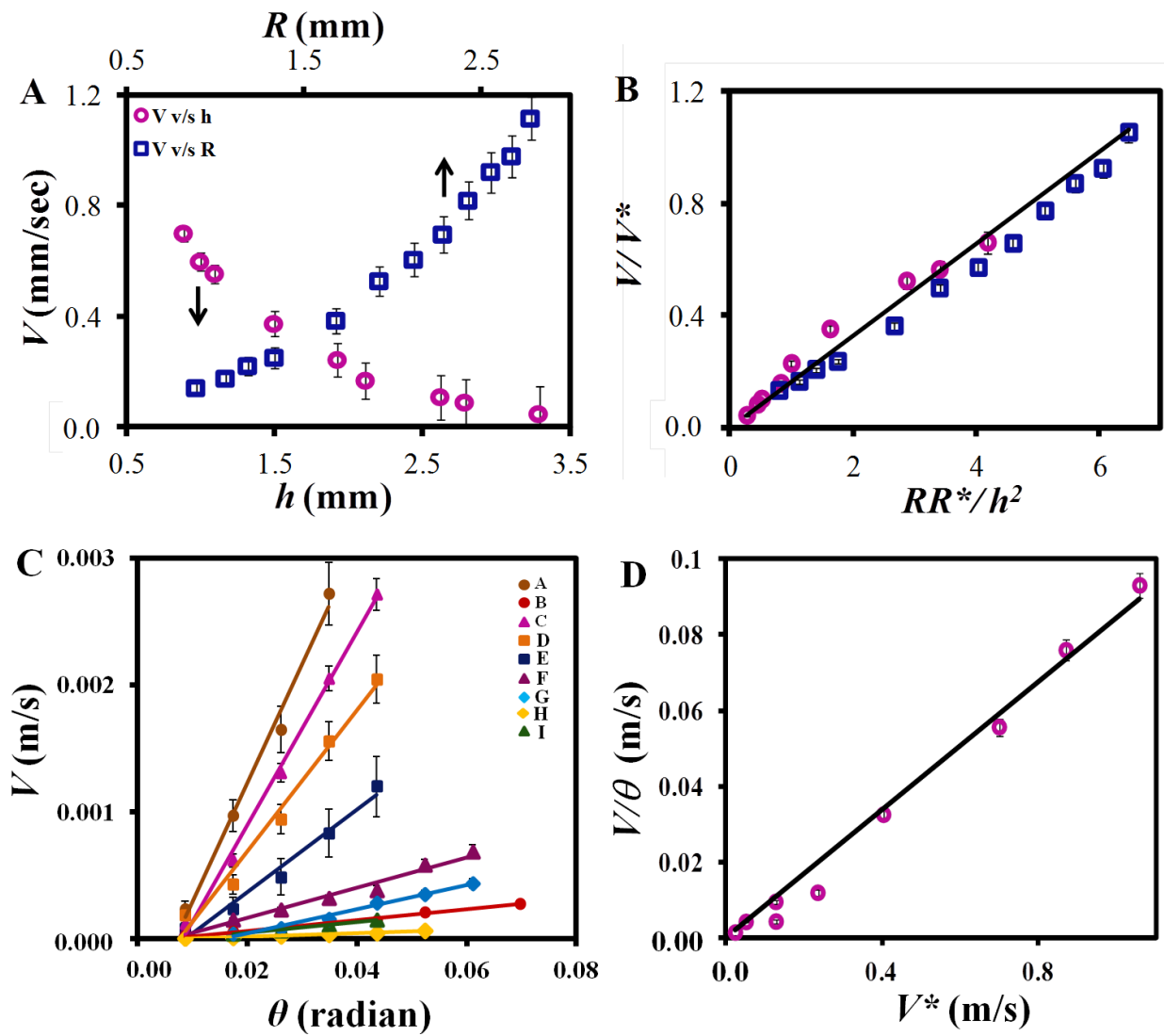

**Figure 5.** A) The velocity of water drop increases non-linearly with its radius (*R*), but decreases non-linearly with the thickness (*h*) of oil above the silicon wafer. These experiments were carried out with AR 20 silicone oil at a wedge angle of 0.66°. B) The data of Figure A are processed by plotting the non-dimensional velocity (*V/V\**) as a function of the non-dimensional geometric factor $RR^*/h^2$, where $R^* = (R(2Rh - h^2))^{1/3}$. C) The velocity of a water drop (volume~11 μl, *h*=1 mm) increases linearly with the wedge angle (θ), but decreases with the viscosity of the liquid. D) The slope of *V*-θ plots of Figure C increases with the capillary velocity (*V\**) of the test liquid.

The data summarized in Figure 5 show that for a drop with a constant volume (11 μL) and *h* (~1 mm), the velocity indeed increases with the wedge angle $\theta \approx |\nabla h|$. Furthermore, a linear relationship is obtained when the slopes of all the $V, \theta$ lines are plotted against the capillary

velocity $V^*$, which is generally consistent with equation 6. There is one notable point in Figure 5, which is that the $V, \theta$ lines do not strictly go through the origin (0, 0) for some of the liquids. This departure occurs at very low wedge angles for which some factors yet unidentified are responsible. In some experiments performed with low viscosity liquids at a wedge angle of $0.25^o$ (or, 0.004 radians), we observed that the drop moves very sluggishly at first, but it slows down and stops eventually. The contact angle of the drop with the solid, however, remains $180^o$ all the way. We suspect that certain impurities in the oil accumulating underneath the drop might contribute to this slowing down. This is indirectly verified by dissolving such an impurity as iodine to the oil, which drastically impedes the motion of the drop at low wedge angles. However, in all cases when this low threshold is overcome, i.e. as long as $\theta > 0.25^o$, water drop always moves.

**Drop manipulation using dynamically evolving topography**

The experiments (Figures 3 to 5) performed with a silicon wafer helped us to identify the range of possibilities that can be exploited to control the motion of a water droplet in a wedge. Based on these results, we are poised to address the next objective of this work that is to modify the landscape of a fluidic platform using a deformable film[50] (figure 6). To this end, we used a polydimethylsiloxane elastomer of modulus ~1 MPa, the surface energy of which is comparable to that of the PDMS grafted silicon wafer. While various silicone oils (Table 1 and Figure 5) could be applied on the surface of such a PDMS film to prevent the adhesion of droplet with the substrate, care had to be taken to ensure that the oil does not swell the PDMS. This requirement eliminates the commonly found silicone and a hydrocarbon based oils. However, a silicone oil in which some of the methyl groups are substituted with phenyl functionality induced little or no swelling of PDMS elastomer. Two such oils, commercially known as AR 20 and a high temperature silicone oil (designated here as HTS), were satisfactory for this purpose, HTS is superior to AR 20 as it induced no swelling of the elastomer. The densities (1.01 to 1.05 gm/cc) of these oils (Table 1) are also not much different from water thereby minimizing the effects of gravity. Thin films of a mixture of two commercial elastomers (Sylgard 184 and Sylgard 186) were cast on a large (diameter 14 cm) polystyrene dish via a spin coating process. The thickness of the resultant film varies somewhat, but it is typically close to 1 mm. While the film is still liquid, small magnetic steel particles (1.56 mm diameter) were placed on the film forming a square matrix array following which the film was cured in an oven to carry out the crosslinking reaction.

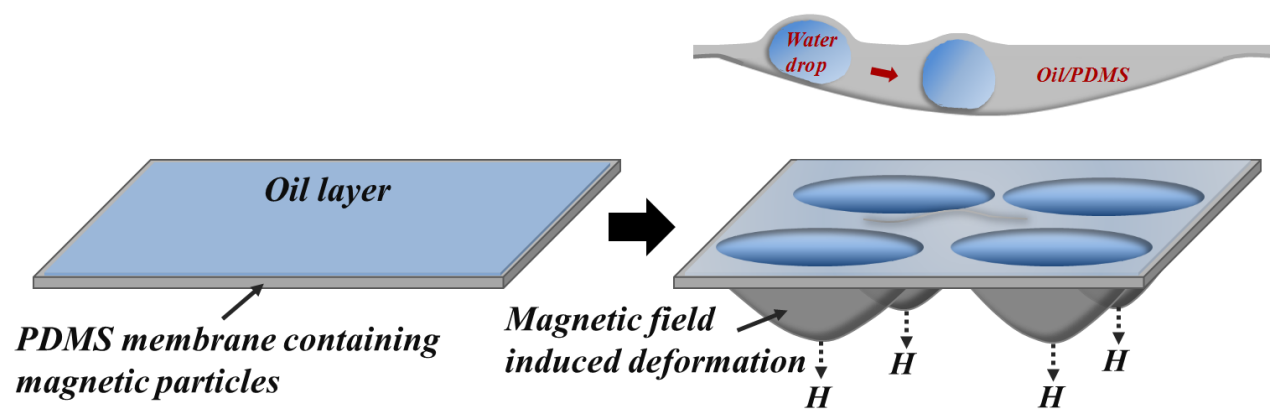

**Figure 6.** Schematic of the modulation of the shape of an oil laden elastomeric membrane of PDMS dispersed with magnetic particles by external electro-magnetic fields that can be simultaneously or sequentially turned on or off. This magnetic field induced deformation of the film alters its landscape with variable curvature that creates wedges thus causing motion of water drops as shown.

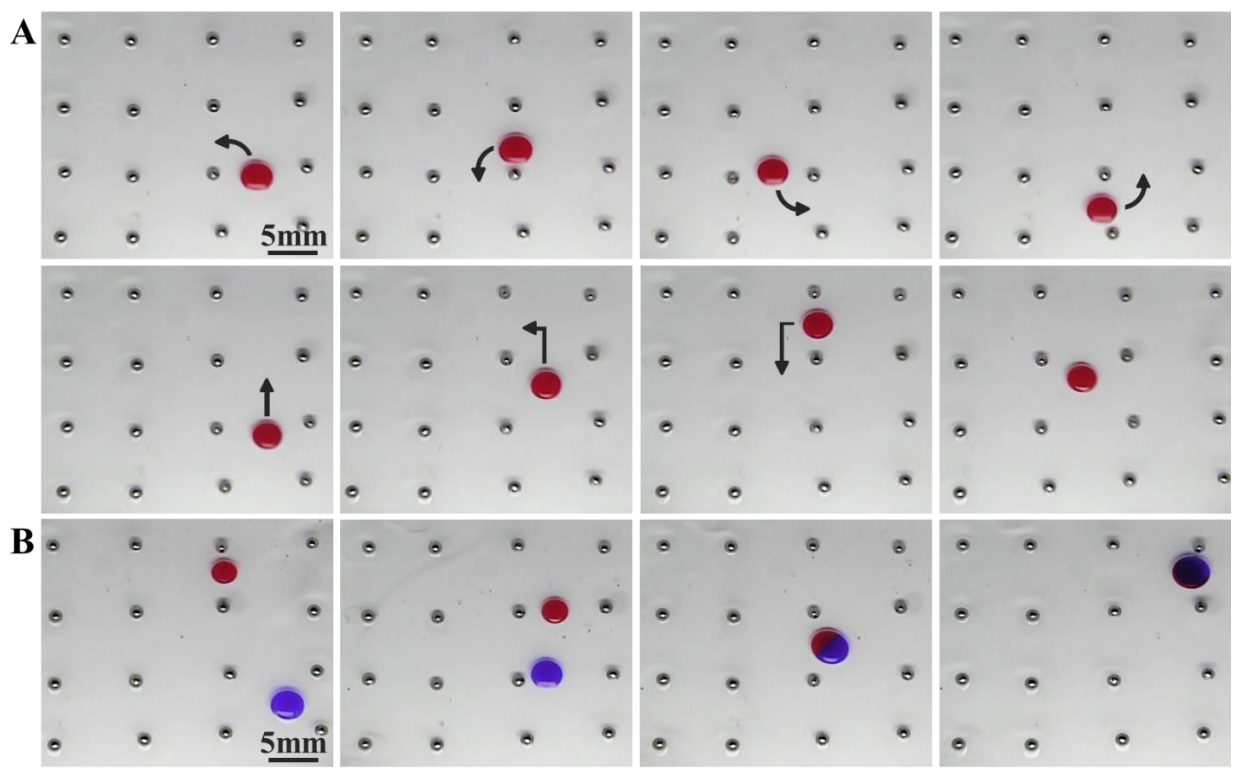

**Figure 7.** Panel A) Small metal spheres are embedded in an elastomeric silicone film that is then coated with a thin film of oil. When a magnetic field is produced close to the metal spheres, the film deforms locally perpendicular to its surface. This creates a capillary wedge or circular symmetry around the particles. A water drop (colored red here) senses this changed landscape and moves towards the minimum of the bell-shaped energy potential. Moving the magnetic field parallel to the substrate, both short and long range motions of the drop, including its rotation and translation can be achieved. Panel B) By controlling the movement of the magnetic field, multiple droplets can be manipulated on the surface. Here two droplets (colored red and blue) are brought close to each other till they fuse. The fused drop can then be transported to another location of the surface. Thickness of oil film 0.4 mm

Once the crosslinking is complete, the metal particle impregnated elastomeric film could be easily lifted off the polystyrene dish and fastened in between two concentric rings (8.5 cm diameter) that generated a certain amount of pretension. Such a film could be vibrated with a gentle tap exhibiting a low frequency (7 Hz) vibration. Using the standard equation of the vibration of the fastened membrane[51], its tension was estimated to be about $T= 1$ N/m. The bending modulus (defined in Supporting Information) of such a film was typically about $D= 2 \times 10^{-5}$ Nm. The combination of this tension and bending modulus yields a decay length $[(D/T)^{0.5}]$ of the deformation of the membrane to be about 4 mm. The metal particles placed at a distance larger than this decay length ensures that they behave more or less independently of each other when one of them is deformed by an external magnetic field. In our experiments the particles were separated by a distance of ~ 10 mm. At this juncture, we remind the readers the results of some previous studies[52, 53] that showed that capillary flow could be enhanced on a deformable wall due to the cooperativity of capillary pressure and the elastic stresses in the film. We feel that this effect should be rather small in our pre-stressed film having an effective elastic tension of 1N/m, which is much larger than any of the surface and interfacial tensions ~ 25 mN/m in our system. Furthermore, as the water drop does not contact the elastomeric film, there is virtually no exchange of mechanical action from one to the other.

After fastening the film in between the concentric rings with the metal particles facing upward, it was leveled and then coated with a thin (0.5 mm) layer of an oil film (AR20 or HTS). The surface of the PDMS elastomeric film was slightly undulated periodically in 2d due to the presence of metal particles, which has the advantage of creating a gently varying energy potential with various minima, where deposited water droplets could be trapped only to be actuated later when a magnetic field is applied beneath the film. The elastomer then deforms locally with a maximum deflection of 3 mm forming a centrosymmetric wedge that creates a Laplace pressure gradient in the drop, causing it to move. By moving the magnetic field spatially, various types of motion can be induced to the droplet. For example, the drop can be made to rotate (Figure 7A; Movie5); it can be translated up to a certain distance on the surface and then altering its direction, or the drop can be made to move sinusoidally mimicking somewhat a reptilian gait (Movie6). When two or multiple droplets are released on the surface, they can be brought into contact with each other till they fuse to become a single droplet (Figure 7B; Movie7).

Having achieved the elementary objectives of the drop motion, we ventured to explore if such a device could be heated locally to carry out a specific temperature dependent chemical transformation. As an illustrative example, we considered a water drop containing a surfactant (Triton X 100) that forms a transparent micellar solution at room temperature. But, when the solution is heated to about 60$^{o}$C, it becomes milky by undergoing a cloud point transition. In order to achieve this objective a laminated heater is placed below the film, where such a transformation is expected to occur. By deforming the film with an external magnetic field in such a way that the deformed region contacts the electrically preheated resistive heater, the drop of the aqueous surfactant solution does indeed exhibit the expected cloud point transition (Figure 8). Such a drop can also be moved out of the heated zone so that it becomes transparent again (Movie8). Such a process can be repeated countlessly without the fear of the underneath elastomeric film being contaminated by the surfactant.

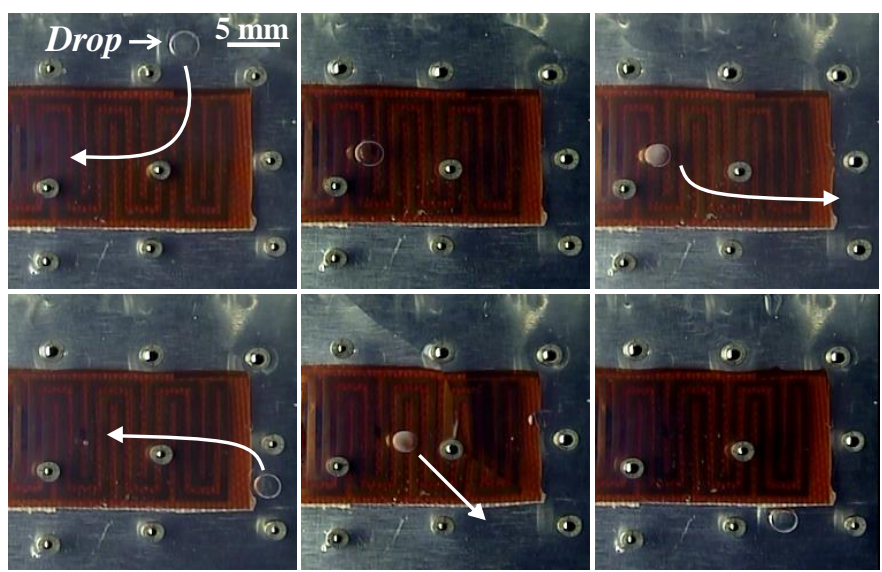

**Figure 8.** The method of drop manipulation as shown in Figure 7 can also be used to carry out temperature sensitive reactions in the droplet. Here a droplet of a micellar solution of a surfactant (Triton X-100) is manipulated on the elastomeric film, by bringing it in contact with the part of the surface that contacts a laminated resistive heater so that its temperature rises to about $60^{\circ}$C. The micellar solution undergoes a cloud point transition at this temperature, causing the drop to be translucent white. When this drop is moved away from the hot zone, it becomes transparent instantaneously as the heat transfer is relatively fast in small drops. There is no evidence of surface contamination as this process of moving the droplet in and out of the hot zone can be repeated countless times.

The operations reported above can be carried out with a simple magnet that can be moved underneath the film. No spatial movement of a magnet is, however, needed in this operation, as the film can be deformed locally by turning electromagnetic fields on and off at different parts underneath the elastomeric film, thereby dynamically altering its landscape. While, metal particles placed in a regular array has the advantage of locally tapping the water droplet that is to be further deformed by external field, it is also possible to carry out this operation by impregnating the elastomeric film with a uniform dispersion of small $80(\pm 10)\mu m$ iron particles.

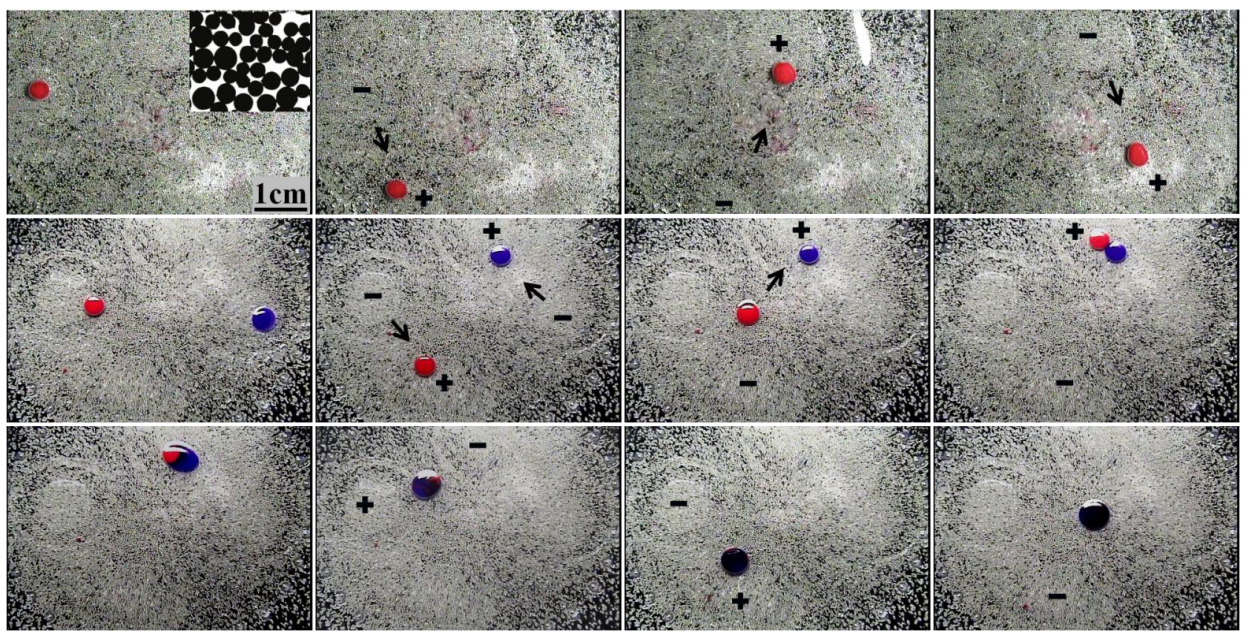

**Figure 9.** Here an elastomeric film impregnated uniformly with small iron particles is used instead of the metal sphere decorated films used above. A microscopic (inset) image of the elastomeric film shows the iron particles, the diameter of which is $80(\pm 10)\mu m$. Versatile movements of a single water droplet (colored red) on this film can be achieved simply by switching the magnetic fields on (+) and off (-) that dynamically modulates the energy landscape of the film. The dynamically evolving energy landscape can also be used to manipulate two

water droplets (colored red and blue), eventually fusing them and moving the fused drop to a different location on the surface. This system should be amenable to logic gates, so that the entire drop dynamics is controlled by a computer.

Here, to begin with, the water drops can be trapped at specific locations of the film by pre-deforming it with spatially inhomogeneous electromagnetic field. By then turning on and off the electromagnetic fields at different locations following a predetermined logic, the motion of a single or multiple drops can be easily manipulated. An example of such an operation is shown in Figure 9, in which the positions of two droplets have been translocated orthogonal to its previous positions and then they were brought into contact to induce fusion (Movie9). The fused drop can afterwards be moved to a different location on the surface, if needed. An image of the iron powder impregnated membrane as seen under a low power microscope (40X) has also been provided in Figure 9.

**CLOSING REMARKS AND FUTURE OUTLOOK**

What we achieved in this work is a controlled actuation of a single or multiple droplets on a surface using a magnetically deformable elasto-capillary device that is simple to fabricate. The droplet manipulations can be carried out repeatedly without the fear of contaminating the elastomeric platform. Various elementary unit operations, such as the translation, rotation, fusion of droplets, and its thermal cycling can be achieved with such a device. Such a drop fluidics is thus poised for carrying out complex biochemical reactions, an examples of which is polymerase chain reaction that require three different distinct steps: (1) denaturation of DNA at 94 °C, primer attachment to the denatured DNA at 55 °C, and then chain extension at 72 °C. From the reaction engineering point of view, what would entail is that a water drop containing the relevant reactants be translocated at the above three thermal zones sequentially, and the cycle be repeated numerous times in order to amplify the number of copies of the original DNA efficiently and rapidly. All these thermal cycling processes should be accomplished with the fluidic concept described here that can be controlled by a pre-programmed computer. An additional advantage of carrying out the reactions with a water drop surrounded by oil film, as described here, is that the evaporation of water would be minimized, thereby making it suitable for carrying out many other biochemical reactions engineering with the aforementioned device. One limiting factor with the reactions carried out with drops is the delay of the time needed to fuse them in a more viscous medium. However, the preliminary studies carried out by us show that the fusion time can be drastically reduced by gently vibrating the elastomeric film that can be carried out with an alternating electromagnetic field superimposed upon a direct field. Future works in our laboratory would target the above objectives.

Before closing the paper, we make a few comments about the theoretical aspects of the work carried out in this paper and what could be done to improve the current analysis. The assumptions that led to the derivation of eq 6 are somewhat simplified. For example, we assumed that the oil film forms a spherical cap above the water droplet. While this assumption is valid when the difference of density between the water drop and the surrounding oil is very small, gravity should play some role when the difference of the density is not negligible. The complete analysis of the drop shape can be carried out by functional minimization of the Lagrangian of the system comprising the gravitational and the interfacial free energies while taking into corrections needed for the thickness dependent van der Waals free energy of the oil film above the drop as it could be thin enough to be influenced by the disjoining pressure acting upon it. While the driving force of the drop propulsion could be obtained following the above analysis, derivation of the viscous resistance that such a drop would experience is more complex that might require a numerical analysis under appropriate level set conditions. All these analyses would also need amendments from the curvature of the elastomeric film that varies in the centrosymmetric wedge following a power law, unlike the analysis carried out in this work. Tools exist to carry out the rigorous analyses as outlined above, but at the cost of losing the simplicity presented in this paper.

Perhaps, a more worthwhile analysis would be to consider the role of the curved wedge[12] on the motion of the droplet as it raises some interesting possibilities. Our preliminary work (Supporting Information) demonstrates that the deformation of the PDMS film conforms to the Föppl von Kármán equation[55] of the deformation of the curved membrane rather well. The curved membrane has the unique feature that the gradient of the surface increases with $h$. It is thus possible that the velocity of a droplet may not decrease (Supporting Information) significantly as it moves towards the wider gap of the wedge, which is not possible with the wedge with a constant gradient.

Another important point to consider is the flow of the oil towards the well created in the elastomeric film as it deforms in response to an external field. The motion of the drop pursuing this external flow partially accounts for the effects reported in Figures 7 and 9 by reducing the drag experienced due to the relative motion of the water drop in the oil itself. This problem is, by no means, an easy one as it would require simultaneous solutions of Navier-Stokes, Föppl von Kármán and capillarity equations; but this is one worth carrying out in future.

Finally, being inspired by the suggestion of one of the referees, we make some comments regarding the scaling of the drop fluidic technology reported here and its relation to the role of gravity. With drops of volumes ranging from 3 $\mu$L to 90 $\mu$L, gravity did not appear to play any significant role as long as the density of the surrounding liquid was close to that of water, which ensures a very large capillary length. As far as very small drops are concerned, we do not see any theoretical limitation of using this technique, as long as the oil film is thin enough and stable. For nanoscopic film, it will be subjected to the so called "disjoining pressure", and of course if there is a gradient of the disjoining pressure, it could be used to induce motion of nano-droplets.

**EXPERIMENTAL DETAILS**

**Materials.** Double side polished silicon wafers of thickness 0.35 mm were obtained from Monsanto. Piranha solution was prepared using Sulphuric acid (98% concentrated, obtained from Fischer chemicals) and Hydrogen peroxide (Fischer Chemicals) in the ratio 7:3 v/v. All Polydimethylsiloxane Trimethylsiloxy Terminated (DMS-T12, DMS-T22 and DMS-T31) test liquids having viscosities 20 cSt, 200 cSt and 1000 cSt respectively were purchased from Gelest. All Phenylmethylsilicones (AR 20 and high temperature silicone oil) having viscosities 23 cSt and 183 cSt respectively were acquired from Sigma Aldrich. Quartz cell (4.5 X 3.9 X 4.8 cm) was purchased from Rame Hart. Steam distilled water of high purity (surface tension 73 mN/m) was used for all the experiments. Iron powder (spherical -40+70 mesh, 99% (metal basis)**,** size $80(\pm 10)\mu m$) and steel balls (Ultra-Hard Wear-Resistant 440C Stainless Steel) of diameters 1.98 mm and 1.56 mm were obtained from Alfa Aesar and McMaster Carr respectively. Electromagnets (diameter 2.54 cm and height 3.175 cm) of maximum pull strength 14.5 Kg were purchased from McMaster Carr. All experiments were recorded using a video camera (Hitachi) connected to a variable focal length microscope (InfiniVar) and analyzed using software Virtual Dub and ImageJ. Specifications of all test liquids are listed in Table1.

**A drop under soft confinement.** A small rectangular (3.8 X 3.4 cm) silicon wafer was cleaned using standard piranha solution and plasma oxidation (Harrick plasma chamber). A thin layer of DMS-T22 was then spread on the surface of the oxidized silicon wafer and it was heated in a hot air oven at 80°C for 24 hours to allow for stable grafting of PDMS on its surface. The unreacted PDMS was then removed by cleaning the wafer in a Soxhlet extractor using chloroform (Fischer Chemicals). In the next step, a quartz cuvette was thoroughly cleaned using chloroform and fixed on a metallic stage (Melles Griot) that could be rotated about two axes with high precision. The treated silicon wafer was placed inside the cuvette on top of a small metallic support, following which, the quartz cell was filled with a test silicone oil such that it attained a controlled thickness above the silicon wafer. The entire set up was fixed atop a vibration isolation table (Micro-g, TMC) to minimize the effects of ground vibration. Once the oil reached a steady level with no air bubbles visible inside, a water drop was gently deposited on the wafer through the oil film using a pipetor (eppendorf) with which the volume of the deposited liquid drop could be accurately controlled.

**Elasto-capillarity induced motion of water droplet sandwiched between an inclined substrate and an air/oil interface.** The set-up used for this experiment was same as that described above**.** Here, the metallic stage supporting the wafer was inclined to a pre-determined angle, following which a drop of water was gently injected either using a syringe (Gilmont$^R$) or a pipetor (eppendorf) on the surface of the oil towards the narrower part of the wedge. As the drop gradually moved towards the thicker part of the wedge, both top and side views of it were video-recorded and subsequently analyzed. This same procedure was iterated for different wedge angles starting from 0.5 to 3.5° for each of the liquids listed in Table1. The viscosities of all the test liquids were measured using traditional Ostwald viscometer (Cannon) using water (viscosity 1cP at 20°c) as the reference liquid. The surface and interfacial tensions of various liquids were measured using a home built drop weight tensiometer.

**Magnetic particle embedded polymeric membrane preparation and its use for drop manipulation.** For all the studies, the elastomeric film was prepared using a mixture of polydimethylsiloxanes (Dow Corning) Sylgard 184 and Sylgard 186 in the ratio of 1:1 w/w. Our previous[54] study showed that such a film is robust against stretching and bending. They could also support significant normal force. Mixtures of these liquid polymers were spun cast on a clean polystyrene petridish (diameter 14cm, obtained from Fischer Chemicals) at 1000 r.p.m for 15 sec in a Photo-Resist Spinner (Headway Research, Inc.). Steel balls (1.56 mm diameter) were placed on the PDMS at a uniform separation of ~ 10 mm from each other in the form of a square grid. The particles penetrated the PDMS film touching the underneath support, while part of it protruded outward above the base of air-film interface. This was then cross-linked in hot air oven at 75°C for 2 hours. After peeling the crosslinked film from the petridish, it was fastened in between two concentric rings (8.5 cm diameter) such that it neither sagged down nor was too stretched to be deformed by magnetic field. This was then placed on a glass plate clamped to an aluminum frame that could be moved translated laterally and tilted. Similar membranes impregnated with iron powder were prepared following the same procedure by sprinkling the iron particles evenly on the surface of the spun cast PDMS film and cured in an oven as described above. A set of four electromagnets connected to a triple output DC power supply (Power Designs INC.) via a home-made on/off switch box was placed below the membrane on a metallic stage. For carrying out temperature sensitive reaction using micellar solution, a laminated resistor heater was placed below the PDMS film on an aluminum plate. After leveling the fastened elastomeric membrane, it was coated with a thin layer of a silicone liquid (phenylmethylsiloxane) ensuring that evenly coats the film. A drop of water was deposited on the fastened membrane through the oil film and it was deformed either by a small magnet or by an electromagnet. When a permanent magnet is used, it had to be spatially moved under the film to cause deformation of the film and the movement of the drop. However, this spatial movement of the magnet is not necessary as the different regions of the film could be deformed or relaxed by switching the electromagnets on and off. While carrying out the temperature sensitive cloud point transition experiment, the temperature (~60°C) of the oil close to the drop was monitored using a digital thermometer (Hewlett Packard).

## SUPPORTING INFORMATION

Text describing additional experiments and movie files. These materials are available free of charge via Internet at http://pubs.acs.org.

## REFERENCES

1. Chaudhury, M. K.; Chakrabarti, A.; Daniel, S. Generation of motion of drops with interfacial contact. *Langmuir* **2015**, 31, 9266-9281.
2. Genzer, J.; Bhat, R. R. Surface-bound soft matter gradients. *Langmuir* **2008**, 24, 2294-2317.

3. Daniel, S.; Chaudhury, M. K.; Chen, J. C. Fast drop movements resulting from the phase change on a gradient surface. *Science* **2001**, 291, 633-636.
4. Chaudhury, M. K.; Whitesides, G. M. How to make water run uphill. *Science* **1992**, 256, 1539-1541.
5. Ouellette, J. A new wave of microfluidic devices. *Industrial Physicist* **2003**, 9, 14-17.
6. Wu, H.; Wheeler, A.; Zare, R. N. Chemical cytometry on a picoliter-scale integrated microfluidic chip. *P. Natl. Acad. Sci. USA* **2004**, 101, 12809-12813.
7. Stone, H. A.; Stroock, A. D.; Ajdari, engineering flows in Small devices: Microfluidics Towards a Lab-on-a-Chip. *Annu. Rev. Fluid Mech*. **2004**, 36, 381-411.
8. Whitesides, G. M. The origins and the future of microfluidics. *Nature* **2006**, 442, 368-373.
9. Wang, B. L.; Ghaderi, A.; Zhou, H.; Agresti, J.; Weitz, D. A.; Fink, G. R.; Stephanopoulos, G. Microfluidic high-throughput culturing of single cells for selection based on extracellular metabolite production or consumption. *Nat. Biotechnol.* **2014**, 32, 473.
10. Greenspan, H. P. On the motion of a small viscous droplet that wets a surface. *J. Fluid Mech.* **1978**, 84, 125-143.
11. Brochard, F. Motions of droplets on solid surfaces induced by chemical or thermal gradients. L*angmuir* **1989**, 5, 432-438.
12. Reyssat, E. Drops and bubbles in wedges. *J. Fluid Mech*. **2014**, 748, 641– 662.
13. Cira, N. J.; Benusiglio, A.; Prakash, M. Vapour-mediated sensing and motility in two-component droplets. *Nature* **2015**, 519, 446-450.
14. Ichimura, K.; Oh, S. K.; Nakagawa, M. Light-driven motion of liquids on a photoresponsive surface. *Science* **2000**, 288, 1624-1626.
15. Petrie, R. J.; Bailey, T.; Gorman, C. B.; & Genzer, J. Fast directed motion of "Fakir" droplets. *Langmuir* **2004**, 20, 9893-9896.
16. Habenicht, A.; Olapinski, M.; Burmeister, F.; Leiderer, P.; Boneberg, J. Jumping nanodroplets. *Science* **2005**, 309, 2043-2045.
17. Narhe, R. D.; Khandkar, M. D.; Shelke, P. B.; Limaye, A. V.; Beysens, D. A. Condensation-induced jumping water drops. *Phys. Rev. E* **2009**, 80, 031604.
18. Boreyko, J. B.; Chen, C.-H. Self-propelled dropwise condensate on superhydrophobic surfaces. *Phys. Rev. Lett.* **2009**, 103**,** 184501.
19. Liu, J.; Guo, H.; Zhang, B.; Qiao, S.; Shao, M.; Zhang, X.; Feng, X.; Li, Q.; Song, Y.; Jiang, L.; Wang, J. Guided Self-Propelled Leaping of Droplets on a Micro-Anisotropic Superhydrophobic Surface. *Angew. Chem. Int. Ed*. **2016**, 55, 4265 – 4269.

20. Li, J.; Hou, Y.; Liu, Y.; Hao, C.; Li, M.; Chaudhury, M. K.; Yao, S.; Wang, Z. Directional transport of high-temperature Janus droplets mediated by structural topography. *Nature Physics* **2016**, doi:10.1038/nphys3643.

21. Pollack, M. G.; Fair, R. B.; Shenderov, A. D. Electrowetting-based actuation of liquid droplets for microfluidic applications. *Appl. Phys. Lett*. **2000**, 77, 1725-1726.

22. Luo, M.; Gupta, R.; Frechette, J. Modulating contact angle hysteresis to direct fluid droplets along a homogenous surface. *ACS Appl. Mater. Interfaces* **2012**, 4, 890-896.

23. Velev, O. D.; Prevo, B. G.; Bhatt, K. H. On-chip manipulation of free droplets. *Nature* **2003**, 426, 515-516.

24. Marcos-Martin, M.; Beysens, D.; Bouchaud, J. P.; Godreche, C.; Yekutieli, I. Self-diffusion and 'visited'surface in the droplet condensation problem (breath figures). *Physica A* **1995**, 214, 396-412.

25. Chaudhury, M. K.; Chakrabarti, A.; Tibrewal, T. Coalescence of drops near a hydrophilic boundary leads to long range directed motion. *Extreme Mechanics Letters*, **2014**, 1, 104-113.

26. Daniel, S.; Chaudhury, M. K. Rectified motion of liquid drops on gradient surfaces induced by vibration. *Langmuir* **2002**, 18, 3404-3407.

27. Daniel, S.; Sircar, S.; Gliem, J.; Chaudhury, M. K. Ratcheting motion of liquid drops on gradient surfaces. *Langmuir* **2004**, 20, 4085-4092.

28. Sandre, O.; Gorre-Talini, L.; Ajdari, A.; Prost, J.; Silberzan, P. Moving droplets on asymmetrically structured surfaces. *Physical Review E*. **1999**, 60, 2964.

29. Reyssat, M.; Pardo, F.; Quéré, D. Drops onto gradients of texture. *Europhys. Lett.* **2009**, 87, 36003.

30. Brunet, P.; Eggers, J.; Deegan, R. D. Vibration-induced climbing of drops. *Phys. Rev. Lett*. **2007**, 99, 144501.

31. Noblin, X.; Kofman, R.; Celestini, F. Ratchetlike motion of a shaken drop. *Phys. Rev. Lett*. **2009**, 102, 194504.

32. Mettu, S.; Chaudhury, M. K. Motion of liquid drops on surfaces induced by asymmetric vibration: role of contact angle hysteresis. *Langmuir* **2011**, 27, 10327-10333.

33. Daniel, S.; Chaudhury, M. K.; De Gennes, P. G. Vibration-actuated drop motion on surfaces for batch microfluidic processes. *Langmuir* **2005**, 21, 4240-4248.

34. Malvadkar, N. A.; Hancock, M. J.; Sekeroglu, K.; Dressick, W. J.; Demirel, M. C. An engineered anisotropic nanofilm with unidirectional wetting properties. *Nat. Mater*. **2010**, 9, 1023–1028.

35. Linke, H.; Alemán, B. J.; Melling, L. D.; Taormina, M. J.; Francis, M. J.; Dow-Hygelund, C. C.; Narayanan, V.; Taylor, R. P.; Stout, A. Self-propelled Leidenfrost droplets. *Phys. Rev. Lett.* **2006**, 96, 154502.

36. Celestini, F.; Frisch, T.; Pomeau, Y. Room temperature water Leidenfrost droplets. *Soft Matter* **2013**, 9, 9535-9538.

37. Abdelaziz, R.; Disci-Zayed, D.; Hedayati, M. K.; Pöhls, J. H.; Zillohu, A. U.; Erkartal, B.; Chakravadhanula V. S. K; Duppel, V.; Kienle, L.; Elbahri, M. Green chemistry and nanofabrication in a levitated Leidenfrost drop. *Nature communications* **2013**, 4, 1-10.

38. Egatz-Gómez, A.; Melle, S.; García, A. A.; Lindsay, S. A.; Márquez, M.; Dominguez-Garcia, P.; Rubio, M. A.; Picraux, S. T.; Taraci, J. L.; Clement, T.; Yang, D. Discrete magnetic microfluidics. *App. Phys. Lett.* **2006**, 89, 34106-34106.

39. Bormashenko, E.; Pogreb, R.; Bormashenko, Y.; Musin, A.; Stein, T. New investigations on ferrofluidics: ferrofluidic marbles and magnetic-field-driven droplets on superhydrophobic surfaces. *Langmuir* **2008,** 24, 12119–12122.

40. Khalil, K. S.; Mahmoudi, S. R.; Abu-dheir, N.; Varanasi, K. K. Active surfaces: Ferrofluid-impregnated surfaces for active manipulation of droplets. *App. Phys. Lett*. **2014**, 105, 041604.

41. Lorenceau, É.; Quéré, D. Drops on a conical wire. *J. Fluid Mech*. **2004**, 510, 29-45.

42. Ju, J.; Xiao, K.; Yao, X.; Bai, H.; Jiang, L. Bioinspired conical copper wire with gradient wettability for continuous and efficient fog collection. *Adv. Mater.* **2013**, 25, 5937-5942.

43. Chaudhury, M. K. Interfacial interaction between low-energy surfaces. *Mater. Sci. Eng*. R. **1996**, 16, 97-159.

44. Johnson, K. L. *Contact mechanics*; Cambridge university press; London, **1985**.

45. Mahadevan, L.; Pomeau, Y. Rolling droplets. *Phys. Fluids* **1999**, 11, 2449-2453.

46. Hadamard, J. S. Mouvement permanent lent d'une sphere liquide et visqueuse dans un liquide visqueux. *Compt. Rend*. **1911**, 152, 1735–1738.

47. Rybczynski, W. Über die fortschreitende Bewegung einer flüssigen Kugel in einem zähen Medium. *Bull. Acad. Sci. Cracow A*. **1911**, 40–46.

48. Kralchevsky, P. A.; Nagayama, K. Capillary interactions between particles bound to interfaces, liquid films and biomembranes. *Adv. Colloid Interfac.* **2000**, 85, 145-192.

49. Chakrabarti A, Chaudhury MK. Attraction of Mesoscale Objects on the Surface of a Thin Elastic Film Supported on a Liquid. *Langmuir* **2015**, 31, 1911–1920.

50. Raphaël, É.; de Gennes, P. G. Wettability controlled by magnetic fields. *C. R. Acad. Sci. IIB* **1997**, 9, 537-543. Note: These authors conjectured a mechanism of modulating the apparent wettability of an elastomeric film by deforming it with an external field. Our problem is very different from this proposition in that the PDMS film is already wetted by the oil. We are, in fact, interested the more important secondary process that is to take advantage of the wedge-like geometry created on an oil-laden elastomeric film in order to break the symmetry of a water drop and then induce its motion via an unbalanced Laplace force.

51. The frequency ($\omega$) of vibration of membrane, in terms of its tension $T$, areal density ($\sigma$) and radius ($r_{max}$) is: $\omega = 0.383\sqrt{(T/\sigma)}/r_{max}$ . See any major textbook on membrane vibration, e.g. Hall, D. E. *Musical Acoustics*, 2nd Ed; Brooks/Cole Publishing, **1991**.

52. Van Honschoten, J. W.; Escalante, M.; Tas, N. R.; Jansen, H. V.; Elwenspoek, M. Elastocapillary filling of deformable nanochannels. *J. Appl. Phys*. **2007**, 101, 094310.

53. Anoop, R.; Sen, A. K. Capillary flow enhancement in rectangular polymer microchannels with a deformable wall. *Phys. Rev. E*. **2015**, 92, 013024.

54. Cambau, T.; Bico, J.; Reyssat, E. Capillary rise between flexible walls. *Europhys. Lett.* **2011**, 96, 24001.

55. Wan, K.-T.; Guo, S.; Dillard, D. A. A theoretical and numerical study of a thin clamped circular film under an external load in the presence of a tensile residual stress. *Thin Solid Films* **2003,** 425, 150-162.